\documentclass[12pt,a4paper]{article}

\usepackage[utf8]{inputenc}
\usepackage[T1]{fontenc}
\usepackage[english]{babel}
\usepackage{amsmath, amssymb}
\usepackage{graphicx}
\usepackage{geometry}
\usepackage{booktabs}
\usepackage{float}
\usepackage{caption}
\usepackage{siunitx}
\usepackage{hyperref}
\usepackage{tabularx}

\begin{document}

\begin{center}
\LARGE \textbf{Kinematics in Context: The Record Jump of Huaso and Larraguibel as a Teaching Resource for Physics}\\[0.3cm]
\large Mauricio Echiburu$^{a,c}$*, José L. Marcos$^{b}$, René Ríos$^{c}$ \& Robinson Moreno Martínez$^{d}$\\[0.2cm]
\small
$^{a}$School of Sciences, Faculty of Life Sciences, Universidad Viña del Mar, Chile\\
$^{b}$Veterinary Medicine, Faculty of Life Sciences, Universidad Viña del Mar, Chile\\
$^{c}$Ph.D. Program in Sciences, mention in Physics, Universidad de Tarapacá, Universidad de La Serena and Universidad de Valparaíso, Chile\\
$^{d}$Neurophysiopathology Laboratory, Center for Neurobiology and Integrative Physiopathology, Institute of Physiology, Faculty of Sciences, Universidad de Valparaíso, Chile\\
\end{center}

\vspace{0.6cm}

\noindent \textbf{Abstract}\\
In 1949, Captain Alberto Larraguibel and his horse Huaso set the world record for equestrian high jump in Viña del Mar, Chile, by clearing a height of 2.47 meters, a mark that remains unbeaten. This work proposes the use of this historical event as a teaching resource for physics, integrating perspectives from biomechanics and veterinary medicine. Based on the analysis of an audiovisual record of the jump, a kinematic model is developed using the \textit{Tracker} software, determining variables such as displacement, velocity, and acceleration of the horse--rider system. The results make it possible to reflect on the biomechanical and physiological factors involved in animal performance, thus linking physics with real biological processes. It is proposed that this interdisciplinary approach, based on authentic cultural and scientific contexts, may promote meaningful learning, motivation, and a more comprehensive understanding of natural phenomena in science education.

\vspace{0.2cm}

\vspace{0.6cm}

\section*{Introduction}

On February 5, 1949, at the Coraceros regiment in the city of Viña del Mar, Chile, Army Captain Alberto Larraguibel and his horse Huaso cleared a height of 2.47 meters in the high-jump event (\textbf{Figure 1}), setting a world record that remains unbeaten to this day. This event constitutes both a sporting and cultural milestone for Chile, and a singular example of the interaction between human skill, animal performance, and the physical principles governing motion.

From a scientific standpoint, equestrian jumping can be divided into five main phases: approach, takeoff, suspension, landing, and departure (Powers \& Harrison, 1999). Each of these stages involves a dynamic balance between muscular force, impulse, and angular momentum, all of which can be quantitatively analyzed using video-analysis tools. Previous studies have shown that jumping performance depends on biomechanical and physiological factors such as the rider's motor coordination, the horse's muscular power, and the elasticity of its tissues (Clayton, 1989; Alexander, 2002; St. George et~al., 2021). This perspective establishes a bridge between physics and the biological sciences, showing how concepts such as impulse, energy, and parabolic trajectory are expressed in complex living systems.

Beyond its historical and biomechanical relevance, the Huaso--Larraguibel jump offers an exceptional educational context for teaching physics. The kinematic analysis of this event using the \textit{Tracker} software allows students to apply concepts of uniformly accelerated motion, energy, and momentum conservation using real and culturally meaningful data. In this sense, the present work proposes a didactic sequence based on the modeling of the equestrian jump, integrating physics, biomechanics, and veterinary science as an interdisciplinary strategy to promote conceptual understanding and motivation in science teaching.

\begin{figure}[H]
\centering
\includegraphics[width=0.8\linewidth]{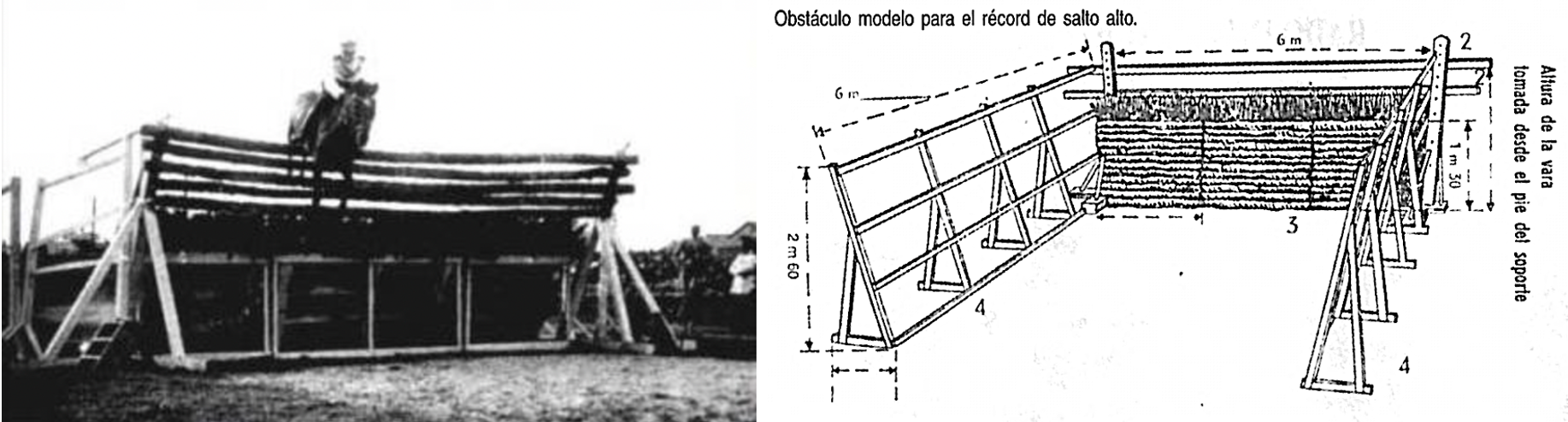}
\caption{Photograph of the Huaso--Larraguibel jump (historical reference).}
\label{fig:1}
\end{figure}

\section*{Biomechanical Description of the Equestrian Jump}

The equestrian jump can be divided into five phases: approach, takeoff, suspension, landing, and departure (Powers \& Harrison, 1999). Each of them presents a specific dynamics that can be interpreted from the principles of classical physics, particularly kinematics and dynamics. The four phases described below are the most relevant for the kinematic analysis of the Huaso--Larraguibel jump.

\textbf{Approach:} In this phase, the horse reaches the speed required to generate the vertical impulse during takeoff. Stride frequency and stride length determine the horizontal velocity, which is limited by the ability of the extensor muscles to generate force (Alexander, 2002). From a physical perspective, the kinetic energy accumulated during this stage is partially transformed into potential energy during the aerial phase of the jump.

\textbf{Takeoff:} This corresponds to the instant when the hind limbs exert the force required to clear the obstacle. This phase is directly associated with the concept of impulse, given by the product of the force applied by the hind legs and the contact time with the ground. The rider's control influences the orientation of the center of mass and the magnitude of the angular momentum (Clayton, 1989).

\textbf{Suspension:} Once airborne, the horse--rider system behaves like a projectile, moving with a velocity and acceleration determined by the initial conditions at takeoff. Conservation of angular momentum allows the system to maintain a stable posture and adjust body rotation by modifying the moment of inertia (Galloux \& Barrey, 1997).

\textbf{Landing:} The main objective is to absorb the impact and recover kinetic balance. In this phase, the ground reaction forces (GRF) act to decelerate the vertical motion while part of the horizontal energy is preserved for the departure phase (Clayton et~al., 1995). This process constitutes a natural example of energy conversion and force control in living systems.

\section*{Kinematic Analysis Methodology and Biomechanical Foundations}

\subsection*{Jump Analysis Using Video and Tracker}

Several audiovisual records of the jump by Huaso and Captain Larraguibel are available on digital platforms, but all of them originate from the same historical video (Youtube, 2011). In this record, the five stages of the equestrian jump can be distinguished: approach, takeoff, suspension, landing, and departure.

For the kinematic analysis, the free software \textit{Tracker} (Tracker, 2022) was used. This software makes it possible to sequence videos and extract motion data frame by frame. Frames 63 to 131 of the record were selected, covering a total of 68 frames with a time interval of \SI{0.02}{s}. This range corresponds to a total time of \SI{1.36}{s}, from the instant when the horse leaves the ground with its hind limbs until several frames after landing.

\begin{figure}[H]
\centering
\includegraphics[width=0.6\linewidth]{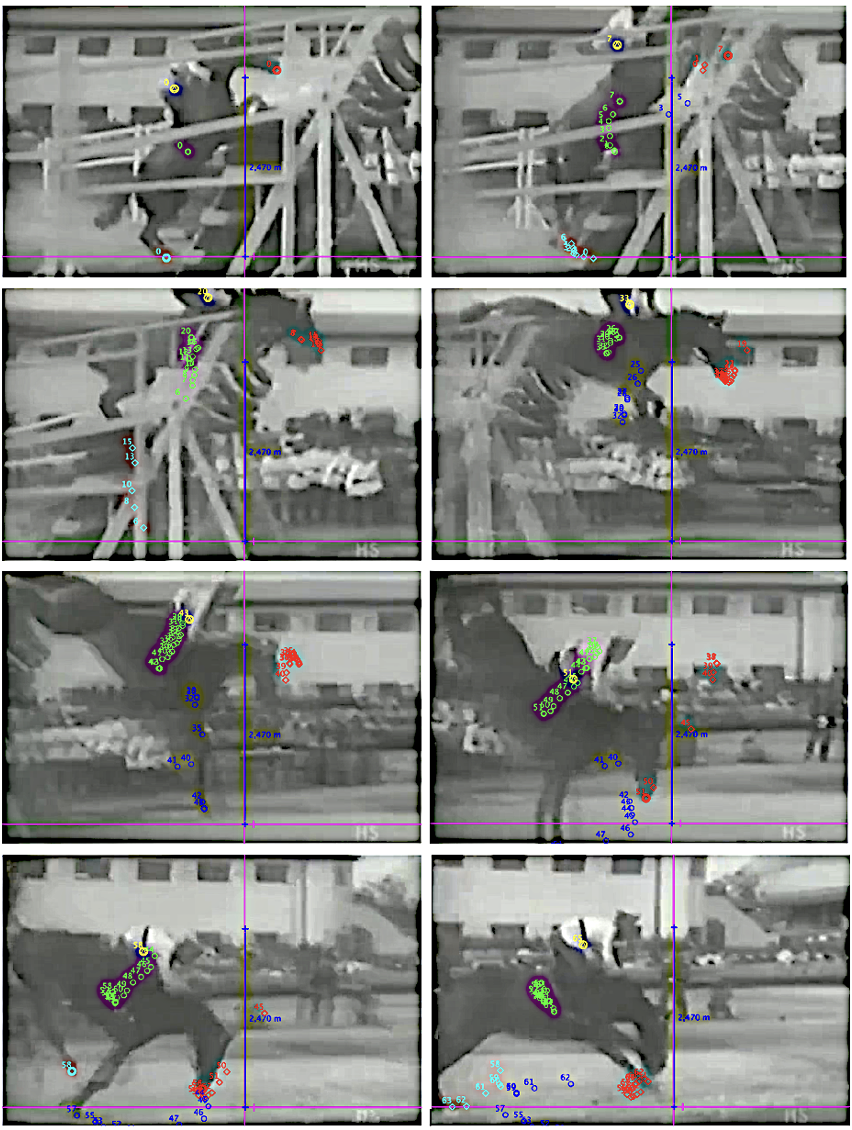}
\caption{Frames extracted from the video and tracking points used in \textit{Tracker}.}
\label{fig:2}
\end{figure}

\textbf{Figure \ref{fig:2}} shows some of the analyzed frames. The perpendicular lines represent the horizontal--vertical coordinate system; the vertical line in a different color corresponds to the \SI{2.47}{m} reference scale; and the numbered points indicate the tracking positions used: horse muzzle (H), forelimb/hand (M), hind limb/leg (P), horse center of mass (CMH), and rider center of mass (CML).

Since the height of the obstacle is known (\SI{2.47}{m}), this value was used as the measurement standard to establish the spatial scale. The recording presents a slight angular sweep and variations in camera height, so the analysis of horizontal displacement was discarded, and the study was focused on vertical motion (the $z$ axis).

From the tracking of the selected points, data tables were constructed to plot the vertical displacement of each point and estimate the impact speed upon reaching the ground. Considering an estimated mass for the horse--rider system, physical quantities such as force, impulse, energy, and power were calculated. This procedure not only made it possible to quantitatively describe the jump, but also to generate a teaching experience applicable to physics education, in which students can analyze a real and culturally meaningful event through concepts of kinematics and dynamics.

\subsection*{Physiological Foundations and Their Relationship to Muscular Performance}

The performance of the equestrian jump depends to a large extent on equine muscle physiology. On average, more than 50\% of the horse's body weight corresponds to skeletal striated muscle tissue, which requires approximately 78\% of total cardiac output. This tissue is designed to sustain high levels of force and speed, supported by highly efficient cardiovascular and respiratory systems.

In intensively trained animals such as Huaso, a higher proportion of fast-twitch type IIa and IIb fibers is expected. Type I fibers are characterized by slow contractions and high endurance, whereas type II fibers display rapid and powerful contractions: type IIa fibers combine high fatigue resistance with abundant mitochondrial content, while type IIb fibers have maximum contraction speed, larger size, and lower myoglobin content. This muscular composition provides the explosive power required for the jump impulse.

\begin{figure}[H]
\centering
\includegraphics[width=0.8\linewidth]{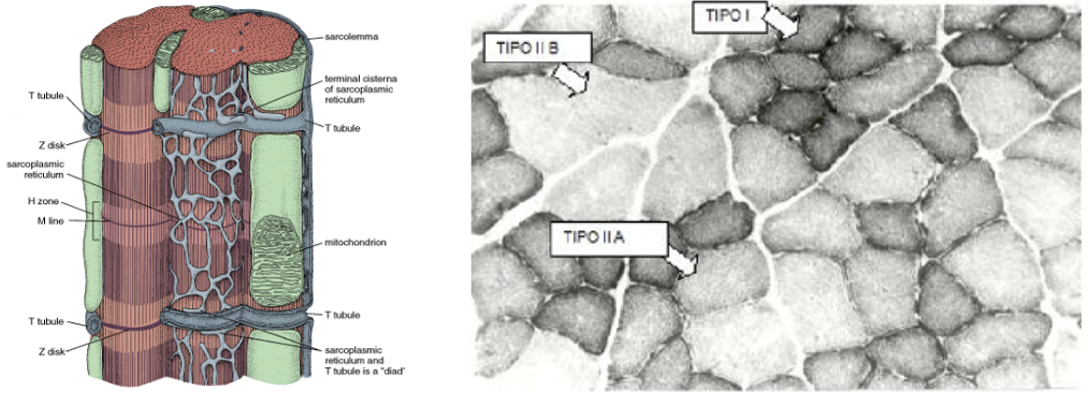}
\caption{(Left) Diagram of the sarcomere and contractile organization of striated muscle. (Right) Types of skeletal muscle fibers in equines (I, IIa, IIb).}
\label{fig:3}
\end{figure}

From a structural perspective, muscle fibers are composed of myofibrils formed by sarcomeres, the functional units responsible for contraction. Within them, the interaction between actin and myosin, mediated by calcium ions and energy in the form of ATP, produces muscle shortening (\textbf{Figure \ref{fig:3}}). This molecular organization does not differ from that of other mammals, but in competitive horses such as Huaso, prolonged training induces adaptations that increase contractile efficiency and the density of myofilaments.

In functional terms, the \textit{Gluteus medius} muscle is particularly relevant, since it directly participates in generating the impulse required for takeoff during the jump. Its rapid and coordinated contraction constitutes an ideal physiological example for discussing the relationship between biological structure, mechanical function, and physical performance.

The analysis of these physiological aspects complements the kinematic description of the Huaso--Larraguibel jump, making it possible to integrate concepts from biology, physics, and veterinary science in the science classroom, while also offering an opportunity to approach science from an interdisciplinary and contextualized perspective.


\section*{Results and Analysis}

From the tracking data obtained with \textit{Tracker}, the vertical displacement of the different points of the horse--rider system was plotted as a function of time. Special attention was given to the rider’s center of mass (CML), since it provides a stable reference for approximating the global motion of the system.

\begin{figure}[H]
\centering
\includegraphics[width=0.7\linewidth]{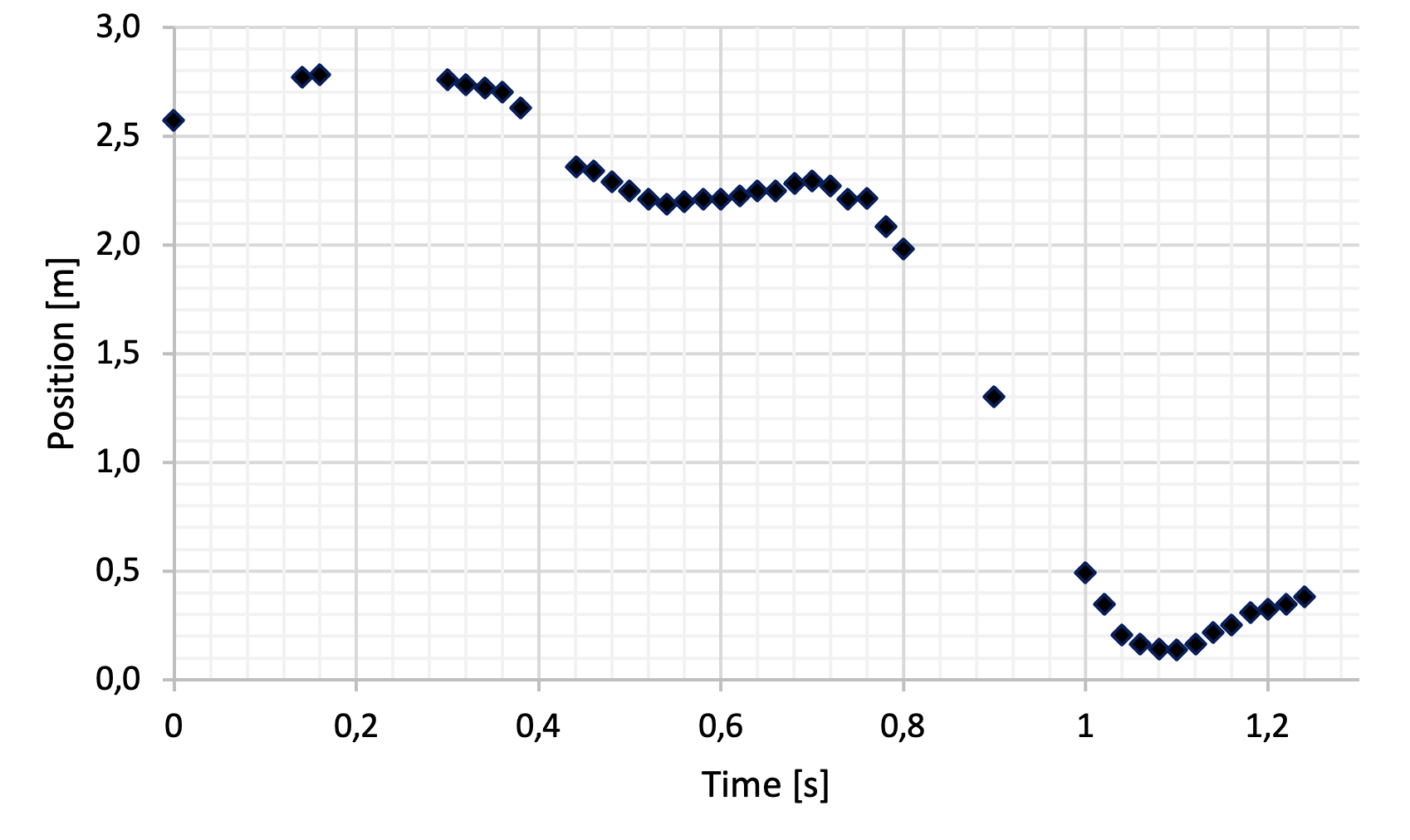}
\caption{Vertical displacement of the rider’s center of mass (CML) and quadratic fit.}
\label{fig:4}
\end{figure}

\textbf{Figure \ref{fig:4}} shows the vertical position of the CML as a function of time, together with a quadratic fit of the form:

\begin{equation}
z(t) = at^2 + bt + c
\end{equation}

The fit yields a value of $a \approx -4.43$, which corresponds to an acceleration of approximately:

\begin{equation}
a = 2a_{\text{fit}} \approx -8.86 \, \text{m/s}^2
\end{equation}

This value differs from the expected gravitational acceleration by approximately 9.6\%, which can be attributed to measurement uncertainty, camera motion, and limitations of the tracking process.

The total flight time obtained from the video is approximately:

\begin{equation}
T_{\text{exp}} \approx 1.36 \, \text{s}
\end{equation}

In contrast, the theoretical prediction obtained from the model is:

\begin{equation}
T_{\text{theo}} \approx 1.42 \, \text{s}
\end{equation}

This represents a relative difference of about 4\%, indicating a good agreement between the theoretical model and the experimental data.

The impact velocity can be estimated from the fitted model, yielding:

\begin{equation}
v \approx 4.20 \, \text{m/s}
\end{equation}

Assuming a total mass of approximately $m \approx 570 \, \text{kg}$ and a contact time of $\Delta t \approx 0.02 \, \text{s}$, the impact force is estimated as:

\begin{equation}
F \approx \frac{m \Delta v}{\Delta t} \approx 1.2 \times 10^5 \, \text{N}
\end{equation}

This corresponds to an acceleration on the order of:

\begin{equation}
a \approx 210 \, \text{m/s}^2 \approx 21g
\end{equation}

The kinetic energy before impact is:

\begin{equation}
K = \frac{1}{2}mv^2 \approx 5.0 \times 10^3 \, \text{J}
\end{equation}

and the associated power during impact is:

\begin{equation}
P \approx \frac{K}{\Delta t} \approx 2.5 \times 10^5 \, \text{W}
\end{equation}


\section*{Role of the Rider in the Jump}

The rider plays a fundamental role in the execution of the equestrian jump, not only by guiding the horse but also by actively influencing the system’s dynamics. During takeoff and suspension, the rider modifies the distribution of mass and the moment of inertia, allowing control over angular stability.

By adopting a forward posture during the suspension phase, the rider reduces the rotational inertia of the system, facilitating a more efficient trajectory over the obstacle. This action is consistent with the conservation of angular momentum, which plays a key role in maintaining stability during flight.

Additionally, the rider’s coordination with the horse allows optimization of the impulse generated during takeoff, directly affecting the maximum height achieved.


\section*{Didactic Proposal}

Based on the analysis presented, a didactic sequence is proposed in which students analyze the Huaso--Larraguibel jump using video tracking tools.

The activity is structured in the following stages:

\begin{itemize}
\item \textbf{Contextualization:} Presentation of the historical event and its relevance.
\item \textbf{Data acquisition:} Use of \textit{Tracker} to obtain position data.
\item \textbf{Modeling:} Application of kinematic equations to describe motion.
\item \textbf{Analysis:} Comparison between experimental data and theoretical predictions.
\item \textbf{Discussion:} Interpretation of results from a physical and biomechanical perspective.
\end{itemize}

This approach promotes active learning and allows students to connect abstract concepts with real-world phenomena.


\section*{Discussion}

The results obtained show that classical mechanics provides a good approximation to describe the motion of the horse--rider system, despite its biological complexity.

The agreement between theoretical and experimental flight times suggests that the projectile model is valid as a first approximation. However, deviations in acceleration and velocity highlight the limitations of the model when applied to real systems.

Factors such as air resistance, internal motion of the system, and measurement uncertainty contribute to these discrepancies. Additionally, the biological nature of the system introduces elements that cannot be fully captured by simple physical models.

From an educational perspective, these limitations provide an opportunity to discuss the scope and validity of physical models, reinforcing the idea that physics is a tool for approximating reality rather than describing it exactly.


\section*{Conclusions}

The analysis of the Huaso--Larraguibel jump demonstrates that real-world phenomena can be effectively used as contexts for teaching physics.

Through video analysis and modeling, it is possible to estimate relevant physical quantities such as acceleration, velocity, force, and energy, and to compare them with theoretical predictions.

This interdisciplinary approach, which integrates physics, biomechanics, and veterinary science, contributes to a more meaningful understanding of natural phenomena and promotes student engagement.

The use of culturally relevant and historically significant events, such as the Huaso jump, represents a valuable strategy for science education.



\end{document}